\documentclass[twocolumn]{article}
\usepackage{preprint}
\usepackage{geometry}  
\usepackage{hyperref}
\hypersetup{colorlinks=true, linkcolor=purple, urlcolor=blue, citecolor=cyan, anchorcolor=black}
\usepackage[numbers,square]{natbib}
\usepackage[utf8]{inputenc}
\usepackage[T1]{fontenc}
\usepackage{tikz}
\usepackage{array}
\usepackage{lineno}
\usepackage{xcolor}
\usepackage{amssymb}
\usepackage{authblk}
\usepackage{caption}
\usepackage{amsmath}
\usepackage{booktabs}
\usepackage{tabularx}

\usepackage{titlesec}
\titlespacing\section{0pt}{12pt plus 3pt minus 3pt}{1pt plus 1pt minus 1pt}
\titlespacing\subsection{0pt}{10pt plus 3pt minus 3pt}{1pt plus 1pt minus 1pt}
\titlespacing\subsubsection{0pt}{8pt plus 3pt minus 3pt}{1pt plus 1pt minus 1pt}
\title{Enhancing Free-hand 3D Photoacoustic and Ultrasound Reconstruction using Deep Learning}

\title{Enhancing Free-hand 3D Photoacoustic and Ultrasound Reconstruction using Deep Learning}
\author[1]{SiYeoul Lee}
\author[1]{SeonHo Kim}
\author[1]{Minkyung Seo}
\author[1]{SeongKyu Park}
\author[1]{Salehin Imrus}
\author[1]{Kambaluru Ashok}
\author[1]{DongEon Lee}
\author[1]{Chunsu Park}
\author[1]{SeonYeong Lee}
\author[1]{Jiye Kim}
\author[3]{Jae-Heung Yoo}
\author[1,2,*]{MinWoo Kim}
\affil[1]{Department of Information Convergence Engineering, Pusan National University, Yangsan, Korea}
\affil[2]{School of Biomedical Convergence Engineering, Pusan National University, Yangsan, Korea}
\affil[3]{Department of Orthopedic Surgery, Busan Medical Center, Busan, Korea}

\begin{document}
\twocolumn[\begin{@twocolumnfalse}
\maketitle

\begin{abstract}
This study introduces a motion-based learning network with a global-local self-attention module (\textbf{MoGLo-Net}) to enhance 3D reconstruction in handheld photoacoustic and ultrasound (PAUS) imaging.
Standard PAUS imaging is often limited by a narrow field of view and the inability to effectively visualize complex 3D structures.
The 3D freehand technique, which aligns sequential 2D images for 3D reconstruction, faces significant challenges in accurate motion estimation without relying on external positional sensors.
MoGLo-Net addresses these limitations through an innovative adaptation of the self-attention mechanism, which effectively exploits the critical regions, such as fully-developed speckle area or high-echogenic tissue area within successive ultrasound images to accurately estimate motion parameters. 
This facilitates the extraction of intricate features from individual frames.
Additionally, we designed a patch-wise correlation operation to generate a correlation volume that is highly correlated with the scanning motion.
A custom loss function was also developed to ensure robust learning with minimized bias, leveraging the characteristics of the motion parameters.
Experimental evaluations demonstrated that MoGLo-Net surpasses current state-of-the-art methods in both quantitative and qualitative performance metrics.
Furthermore, we expanded the application of 3D reconstruction technology beyond simple B-mode ultrasound volumes to incorporate Doppler ultrasound and photoacoustic imaging, enabling 3D visualization of vasculature.
The source code for this study is publicly available at: \underline{\url{https://github.com/guhong3648/US3D}}
\end{abstract}


\vspace{0.5cm}

\end{@twocolumnfalse}]

\section{Introduction}
Diagnostic ultrasound (US) is a widely employed modality 
for examining various organs and tissues, owing to its real-time imaging capabilites.
Particularly, it serves as a preferred tool for guiding medical procedures 
such as biopsies and injections \cite{ho_2021_US_basic}.
In addition, US plays a pivotal role in dynamically monitoring the vascular system
through the Doppler effect \cite{evans_2011_US_doppler_basic}.
Meanwhile, the integration of photoacoustic (PA) imaging with conventional US imaging 
(PAUS imaging) has been extensively studied because of its promising clinical potential. 
This integration aims to enhance the existing benefits of ultrasound 
while introducing novel capabilities for both interventional and functional imaging 
\cite{lee_2023_PA_panoramic}.

In the PAUS system, a handheld transducer is responsible 
for emitting US or/and laser pulses and receiving resulting acoustic wave signals. 
While this configuration allows users the flexibility to manually scan a region of interest, 
it is accompanied by certain limitations. 
Notably, the field of view is restricted, 
providing only a narrow 2D cross-section of the image area, 
which hampers the understanding of the topological 3D structures of the target. 
Although some transducers are specifically designed for 3D imaging, 
their physical dimensions often render them impractical 
for handheld use \cite{mozaffari_2017_US_freehand_review}.

An alternative approach is the 3D freehand method, 
which involves the sequential alignment of 2D image sections acquired 
from freehand sweeps with a standard transducer. 
Users can playback a series of 2D US image frames as a video through a CINE loop, 
but the visualization of the 3D rendering volume or structure of the target 
through accumulating frames is more straightforward. 
This method proves ideal in clinical practice when tissue motion is static. 
However, a primary technical challenge of this approach lies in estimating sweeping motion 
for correct 3D reconstruction. 
Although an external positional device can be attached to the transducer, 
it results in making the transducer bulkier and often provides inaccurate measurements 
in a clinical environment due to various optical or electrical disturbances
\cite{sorriento_2019_sensor_limitations}.

Several studies have explored the estimation of scan trajectories directly from 2D US images, eliminating the need for external sensors \cite{chen_1997_SD_1,tuthill_1998_SD_2,gee_2006_SD_3}. The key idea is to exploit and track tissue speckle patterns in US B-mode images. Recent approaches have employed deep learning (DL) frameworks to better leverage speckle features for predicting scan motion, typically defined by six parameters. Since the pioneering work of Prevost et al. \cite{prevost_2018_CNN}, which introduced the first DL model for this task, subsequent studies have proposed advanced models \cite{guo_2020_DCL, luo_2021_LSTM, ning_2022_transformer, miura_2021_Absolute_pose, luo_2023_RecO} to improve stability and accuracy.
However, their overall estimation accuracies were still insufficient for clinical practice, 
and they were rarely validated for long elevational scan trajectories, 
which are common in clinical practice.

The principal contribution of this study lies in 
proposing a novel deep learning method to elevate visualization capabilities of 3D US and PAUS system.
The structure is composed of a ResNet-driven encoder \cite{li_2023_EfficientNet} 
and an estimator with a Long Short-Term Memory (LSTM) block 
to extract features from given B-mode frames 
and estimate the motion vectors between  consecutive frames, 
leveraging long-term memory.
One part of the encoder consists of special blocks that can directly access 
the correlation between encoded feature maps from adjacent frames. 
Additionally, we designed task-specific novel global-local attention module, 
which effectively highlights the critical local regions for motion estimation, 
such as fully-developed speckle areas or high-echogenic tissue areas. 
Moreover, our customized loss function ensures that the network robustly learns motion 
without a significant distortion in specific parameter estimation.

The model was rigorously evaluated using both our proprietary dataset and a publicly available dataset.
Utilizing a programmable ultrasound system enabled us to assess the model’s performance 
across a variety of conditions, such as reductions in B-mode image speckle 
or the inclusion of more unprocessed data as input. 
Comparing deep learning methods is inherently difficult due to the fact 
that each model is typically designed and optimized for a specific acquisition dataset. 
To mitigate this issue, we also employed an publicly open dataset \cite{li_2023_EfficientNet}, obtained under entirely different settings from our own. By training and testing our model on open dataset, we aimed to demonstrate that our framework is generalizable and not overly dependent on the dataset we specifically acquired.

Another significant contribution involves highlighting the versatility of the panorama technique 
in clinical applications, extending beyond the mere compilation of B-mode images. 
Through the combination of US and laser transmission sequencing, 
hand-held free-scan PAUS system can obtain PA data as well as US data at once. 
Using positioning contents, we demonstrate that 3D reconstruction of vessels is available 
from either US power Doppler mode (PD-mode) data or PA data. 
To the best of our knowledge, this is the first reported application of this approach. 
These advancements have the potential to aid specialists in diagnosing diseases 
or precisely locating targets during the intervention process.


\section{Related Works}
\label{sec:Related_Works}

Traditional approaches exploited tissue speckle patterns in US B-mode images, 
as speckle content between successive frames tends to be preserved, 
even in out-of-plane  motions. 
Since the pioneering work by Trahey et al. \cite{trahey_1986_SD_0}, 
several studies have focused on estimating scan motion by analyzing the correlation 
between adjacent frames \cite{chen_1997_SD_1, tuthill_1998_SD_2}. 
Gee et al. \cite{gee_2006_SD_3} proposed an adaptive speckle decorrelation technique, 
leading to improved performance. 
These studies have demonstrated the feasibility of sensorless free-hand US 
through gradual improvements in various factors such as estimation accuracy, 
generalizability, and less constrained scan protocols.

With the rapid advancements in DL, 
Prevost et al. \cite{prevost_2018_CNN} were the first to attempt 3D US volume reconstruction 
using a convolutional neural network (CNN).
Guo et al. \cite{guo_2020_DCL} introduced a deep contextual learning network (DCL-Net), 
utilizing 3D convolution to exploit the sequential context information in US scan frames, 
along with an innovative loss function based on correlation values. 
Building on this, Guo et al. \cite{guo_2022_DC2} expanded their previous work 
by developing a deep contextual-contrastive network (DC\textsuperscript{2}-Net), 
which applied a margin triplet loss for contrastive learning in a regression task.

Around the same time, 
Luo et al. \cite{luo_2021_LSTM} made key contributions to free-hand US imaging 
by proposing an RNN-based model with a novel self-supervised and adversarial learning strategy. 
This approach enabled plausible visual reconstructions, 
even in more challenging scanning scenarios. 
Luo et al. continued their active contributions with the development of MoNet, 
a motion network incorporating an inertial measurement unit (IMU) sensor, 
a lightweight sensor for capturing acceleration data \cite{luo_2022_MoNet}. 
Their multi-branch DL architecture leveraged both US images and IMU sensor data for improved performance. 
In later work, Luo et al. utilized multiple IMU sensors to further enhance reconstruction accuracy \cite{luo_2023_OSCNet}.

Recently, Luo et al. \cite{luo_2023_RecO} 
introduced an online learning reconstruction framework (RecON), 
imposing constraints such as motion-weighted training loss, frame-level contextual consistency, 
and path-label similarity, which significantly improved the accuracy of motion estimation 
in complex scan motions. 
Other recent DL approaches have adapted popular models for US imaging. 
Miura et al. \cite{miura_2021_Absolute_pose} proposed a sequential CNN-RNN structure 
for both relative and absolute pose estimation, 
demonstrating the efficacy of relative pose integration via RNNs. 
Inspired by pyramid warping techniques \cite{tehrani_2020_PWC_Net}, 
Xie et al. \cite{xie_2021_PW_Net} developed a network 
that extracts multi-scale features from US frames 
to better capture low-frequency B-mode information. 
With the increasing popularity of transformers, 
Ning et al. \cite{ning_2022_transformer} applied a transformer architecture 
to combine local and long-range information from a CNN-based backbone encoder and IMU sensors. 
Li et al. \cite{li_2023_EfficientNet} further identified long-term dependencies 
between sequential US frames and the influence of anatomical or scan protocol factors.

Despite these advancements, achieving the level of precision required for clinical applications remains a challenge. Many existing methods have not been validated on extensive, real-world datasets, nor have they adequately addressed the effects of image processing techniques, such as speckle reduction, which can distort critical information in B-mode images. Furthermore, there has been limited exploration of adapting these DL-based methods for other imaging modes, such as power Doppler (PD) and photoacoustic (PA) imaging.

The potential of DL-based scan motion tracking systems to enable vascular visualization over large regions is significant. By integrating optimized ultrasound and laser sequencing, raw data reconstruction, and post-processing techniques, these systems can extend their utility to PA imaging and PD-mode US imaging. Such advancements promise to open new avenues for clinical applications, including more accurate visualization of vascular structures and enhanced interventional guidance.

\section{Methods}
\label{sec:Methods}
The primary aim of this study is to reconstruct a 3D US volume from sequential 2D B-mode frames 
captured by sweeping a standard  transducer over time.
Accurate assembly of the volume hinges on the precise estimation of each frame’s position,
that is, the accumulation of transducer’s motion.
In hand-held PAUS imaging, the real-time B-mode images serve as the foundation.
Due to their distinct anatomical layers and tissue speckle patterns, 
they can aid in tracking the motion.

\subsection{Problem Definition}
Assuming a scan sequence comprises a total of $N$ B-mode frames, 
the absolute position vector $\boldsymbol{\theta}_{i}$ 
of the $i$th B-mode frame $\mathbf{B}_{i}\in\mathbb{R}^{256\times256}$ 
(or transducer at the time) 
is represented by 6-degree of freedom (DoF) vector, 
defining the 2D Euclidean plane for each image frame within the 3D space:
\begin{align}
        \label{eq:absolute_pose_vector}
        \boldsymbol{\theta}_{i} = 
        \left[\theta_i^1, \theta_i^2, \theta_i^3, 
        \theta_i^4, \theta_i^5, \theta_i^6\right]^\top\in\mathbb{R}^6,
\end{align}
where $\left[\theta_i^1, \theta_i^2, \theta_i^3\right]^T$ 
and $\left[\theta_i^4, \theta_i^5, \theta_i^6\right]^T$ 
denote the 3D Euclidean coordinates and the Euler angles. 
Here, $\theta^1, \theta^2, \theta^3$ correspond to the positions 
along the axial, lateral, and elevational axes, 
while $\theta^4, \theta^5, \theta^6$ represent the rotations 
around the pitch, yaw, and roll axes, respectively. 
For $i=0$, $\boldsymbol{\theta}_0$ is initialized as $\mathbf{0}$.

The vector $\theta_i$ can be converted into the homogeneous transformation matrix $\mathbf{T}_{i} = [\mathbf{R}_{i}, \mathbf{t}_{i};\mathbf{0}^\top, 1]$ where $\mathbf{R}_{i} \in\mathbb{R}^{3\times3}$ and $\mathbf{t_i} \in\mathbb{R}^{3}$ indicates the rotation matrix and translation vector, respectively. 
Then, the relative transformation $\Delta\mathbf{T}_{i}$ 
between adjacent frames $(\mathbf{B}_{i}, \mathbf{B}_{i+1})$ can be obtained from 
the transformation matrices ($\mathbf{T}_{i}$, $\mathbf{T}_{i+1}$) as $\Delta\mathbf{T}_{i} = \mathbf{T}_{i+1}\mathbf{T}_{i}^{-1}$.
The cumulative product of the relative transformation matrices 
results in the absolute transformation matrix as
\begin{align}
        \label{eq:cumprod_transformation_matrix}
        \mathbf{T}_{n+1} = 
        \prod_{i=0}^{n} \Delta\mathbf{T}_{i}\mathbf{T}_{0},
\end{align}
where $\mathbf{T}_{0} = \mathbf{I}$. 
Finally, the relative position vector $\Delta\boldsymbol{\theta}_{i}$ is extracted 
from $\Delta\mathbf{T}_{i}$ for use in supervised learning.

Our goal is to estimate the relative motion $\Delta\boldsymbol{\theta}_{i}$ between two consecutive frames $(\mathbf{B}_{i}, \mathbf{B}_{i+1})$ using information from the two frames as well as preceding frames. This can be formally expressed as
\begin{align}
        \label{eq:model_goal}
        f(\mathbf{B}_{i}, \mathbf{B}_{i-1},\mathbf{B}_{i-2},\dots) = \Delta\hat{\boldsymbol{\theta}}_{i},
\end{align}
where $f$ is the DL model, and $\Delta\hat{\boldsymbol{\theta}}_{i}$ is the estimated relative motion. 
The estimate $\hat{\boldsymbol{\theta}_{i}}$ is then used to derive the absolute transformation matrix $\hat{\mathbf{T}_i}$, which is employed to map voxel positions and reconstruct the 3D volume. 

\begin{figure*}[t]
        \captionsetup{labelfont=bf}
        \centerline{\includegraphics[width=\textwidth]{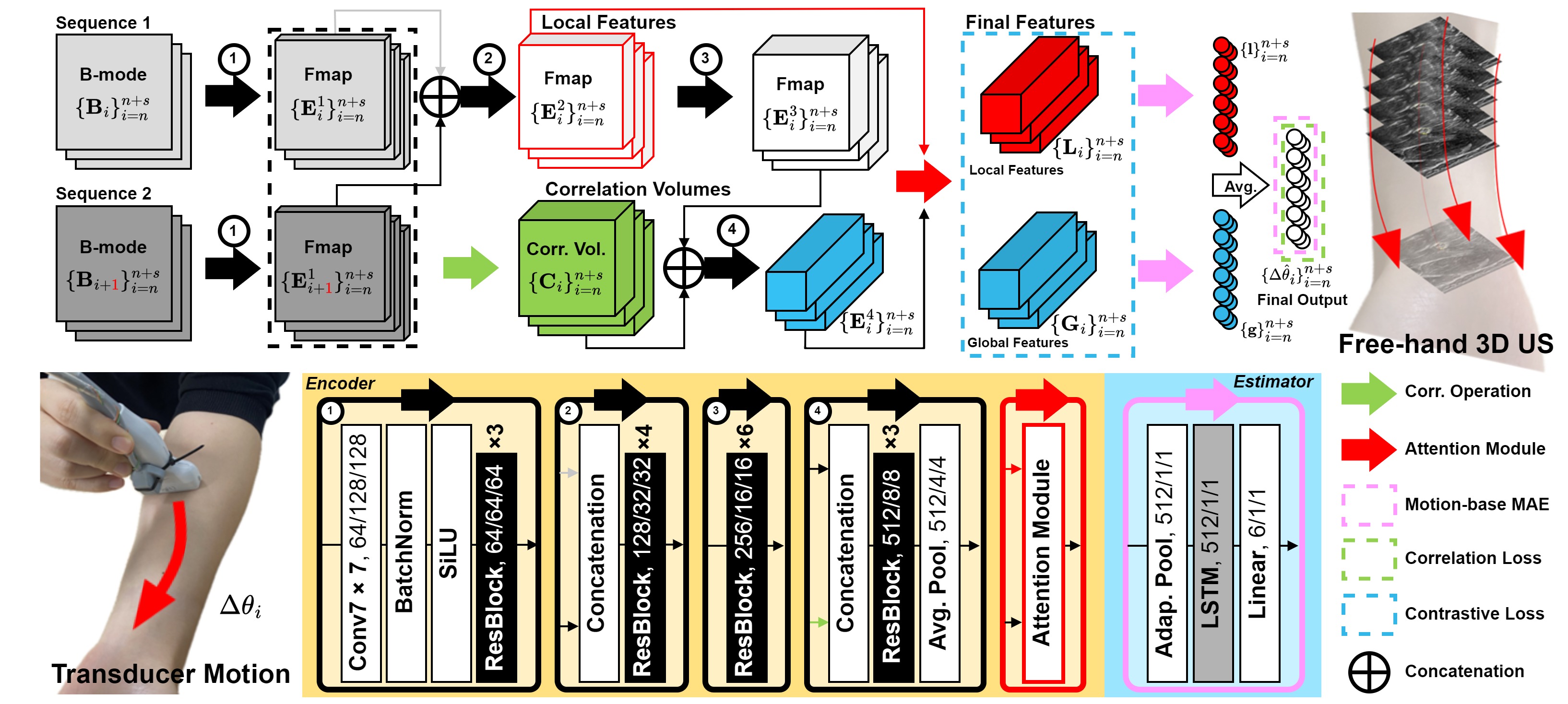}}
        \caption{Overview of our motion-based learning network 
        with a global-local self attention module (MoGLo-Net) structure. 
        Trainable components are denoted by color-filled arrows, 
        signifying the neural networks. 
        Rectangular or cubic shapes represent 2D images or 3D tensors, respectively. 
        The model processes two B-mode sequences 
        and outputs estimates of relative motion vectors $\hat{\Delta\boldsymbol{\theta}_i}$.
        Vectors or feature maps within dotted boxes contribute to the loss function, 
        while the final estimates facilitate the assembly of 2D images into a 3D volume.}
        \label{fig:DL_model}
\end{figure*}

\subsection{Deep Learning Framework}
\begin{figure}[ht]
        \captionsetup{labelfont=bf}
        \centerline{\includegraphics[width=\columnwidth]{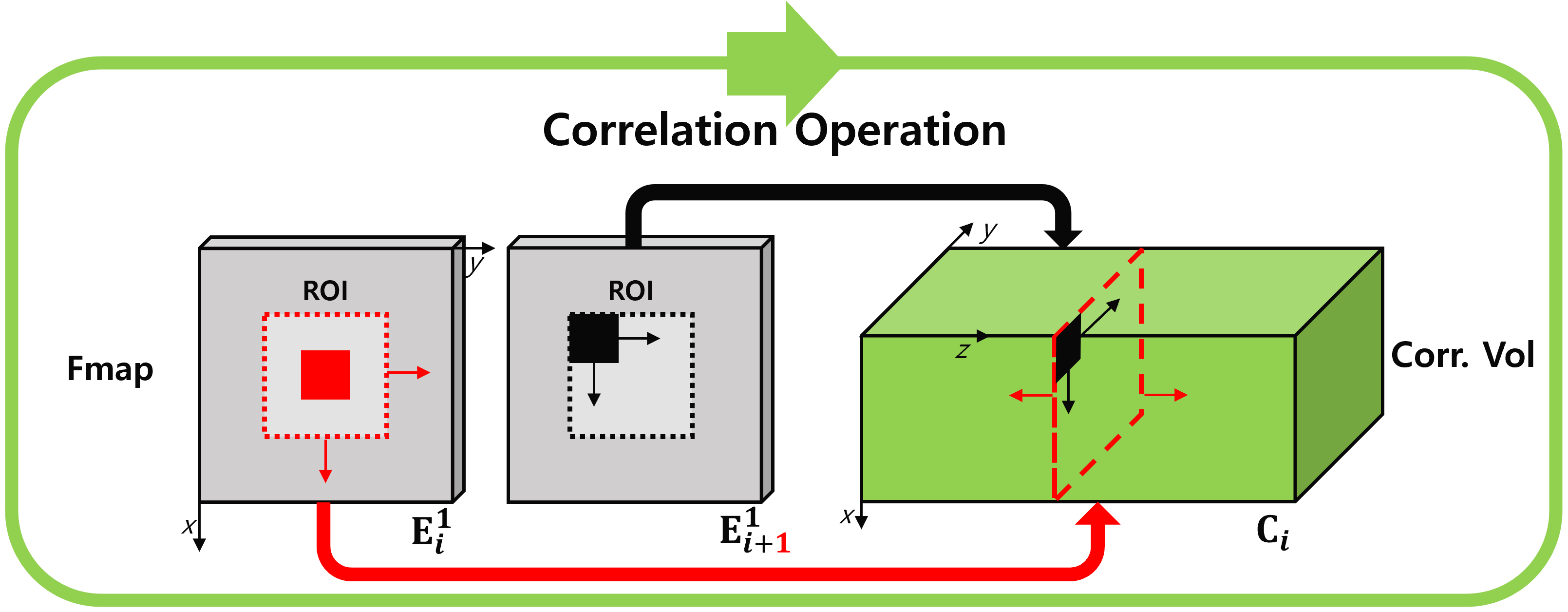}}
       \caption{Correlation operation. It generates the correlation volume $\mathbf{C}_i$ from two feature maps $(\mathbf{E}_{i}^1, \mathbf{E}_{i+1}^1)$. The dotted box in each map represents the spatial region of interest (RoI), and the filled box represents a 3D patch spanning all channels but covering only part of the spatial RoI. The red patch remains fixed at the center of the RoI, while the black patch moves across the RoI. All possible correlations between the two patches are stored in a 2D array (red dotted box) within the volume. By moving both RoIs across the feature maps, these arrays are stacked to generate the full 3D correlation volume. }
\vspace{-10pt}
        \label{fig:Corr}
\end{figure}

As shown in Fig.~\ref{fig:DL_model}, we developed a motion-based learning network, called \textbf{MoGLo-Net} (motion-based learning network with a global-local self-attention module), to estimate the relative scan motion $\hat{\Delta\boldsymbol{\theta}_i}$ from B-mode sequences.
 
\subsubsection{Data Preparation} 
Prior to feeding data into the model, two B-mode sequences were generated. The first sequence, denoted as $\{ \mathbf{B}_i \}_{i=n}^{n+s}$, consists of frames from $\mathbf{B}_n$ to $\mathbf{B}_{n+s}$. The second sequence, $\{ \mathbf{B}_{i+1} \}_{i=n}^{n+s}$, consists of frames from $\mathbf{B}_{n+1}$ to $\mathbf{B}_{n+s+1}$. These sequences are processed in parallel using a ResNet-based Encoder Block 1 \cite{he_2016_ResNet}, ensuring consistent refinement of features. The resulting feature maps are represented as $\{ (\mathbf{E}_{i}^1, \mathbf{E}_{i+1}^1) \}_{i=n}^{n+s}$.  

\subsubsection{Feature Extraction} 
From these features, correlation volumes $\{ \mathbf{C}_i \}_{i=n}^{n+s}$ are extracted to capture relationships between successive frames. The concatenated features from the two sequences are then fed into Encoder Block 2, producing refined feature maps, denoted as $\{ \mathbf{E}_{i}^2 \}_{i=n}^{n+s}$. These features are further processed by Encoder Block 3, which outputs $\{ \mathbf{E}_{i}^3 \}_{i=n}^{n+s}$. Next, the concatenation of the correlation volumes $\{ \mathbf{C}_i \}_{i=n}^{n+s}$ and the features from Encoder Block 3, $\{ \mathbf{E}_{i}^3 \}_{i=n}^{n+s}$, is passed through Encoder Block 4, yielding $\{ \mathbf{E}_{i}^4 \}_{i=n}^{n+s}$.  

\subsubsection{Global-local Attention} 
To refine both global and local contextual information, we employed not only conventional attention but also self-attention mechanism. 
These combined attention mechanisms recalibrate both global and local features while effectively highlighting critical regions for motion estimation within the local features. 
The resulting final features, denoted as $\{\mathbf{G}_i\}_{i=n}^{n+s}$ and $\{\mathbf{L}_i\}_{i=n}^{n+s}$, which are derived from the global and local features, respectively. 
The final features are then contrasted using a triplet loss that leverages motion vectors to improve the model's ability to differentiate subtle variation in motion.

\subsubsection{Motion Estimation} 
For motion estimation, the model employs two RNN-based motion estimators to predict six-dimensional relative motion vectors. The motion, denoted as \(\{ \hat{\Delta\mathbf{g}_i} \}\), is derived from the global features \(\{\mathbf{G}_i\}\), while the motion, denoted as \(\{ \hat{\Delta\mathbf{l}_i} \}\), is derived from the local features \(\{\mathbf{L}_i\}\). The final motion predictions, \(\{ \hat{\Delta\boldsymbol{\theta}_i} \}_{i=1}^{N}\), are obtained by combining these motion vectors. The predictions are supervised using a motion-based mean absolute error (MMAE) and a correlation loss to ensure accurate motion estimation.

\subsubsection{Testing Phase} 
The predicted relative motions $\{ \hat{\Delta\boldsymbol{\theta}_i} \}_{i=1}^{N}$ are transformed into relative transformations, denoted as $\{ \hat{\Delta\mathbf{T}_i} \}_{i=1}^{N}$. Using the cumulative product of these transformations, as defined in Equation~(\ref{eq:cumprod_transformation_matrix}), the absolute transformations $\{ \hat{\mathbf{T}_i} \}_{i=1}^{N}$ are obtained. These absolute transformations are then used to reconstruct the 3D US volumes.  

\subsection{Correlation Operation}  
We devised a patch-wise correlation operation to extract the relationship between adjacent frames, inspired by speckle decorrelation techniques \cite{chen_1997_SD_1, tuthill_1998_SD_2, gee_2006_SD_3}. This operation is performed on the feature maps $(\mathbf{E}_{i}^1, \mathbf{E}_{i+1}^1)$, which are initially refined from the B-mode sequences using Encoder Block 1 (Fig.~\ref{fig:DL_model}).  

As shown in Fig.~\ref{fig:Corr}, the first step in this operation is to define two regions of interest (RoIs), represented by dotted squares, at the same locations in both feature maps $(\mathbf{E}_{i}^1, \mathbf{E}_{i+1}^1)$. Within these RoIs, pairs of patches are extracted. While the patch in $\mathbf{E}_{i}^1$ (red filled square) remains stationary at the center of the RoI, the patch in $\mathbf{E}_{i+1}^1$ (black filled square) is moved across the RoI. A correlation array is computed from all possible pairs of patches within the RoIs. By stacking these correlation arrays from multiple RoIs, the correlation volume $ \mathbf{C}_i $ is generated.  

This correlation volume provides valuable motion-related information, enabling the model to estimate motion more precisely. For instance, if there is no motion (i.e., a stationary frame), the correlation volume ideally contains uniform values. In the case of pure elevational motion, the correlation values generally decrease overall, with the largest value remaining at the center of the correlation arrays when the positions of the two patches coincide. Conversely, during lateral motion, the locations of the largest correlation values shift according to the motion direction.  

\subsection{Global-Local Attention}
\begin{figure}[!t]
       \captionsetup{labelfont=bf}
        \centerline{\includegraphics[width=\columnwidth]{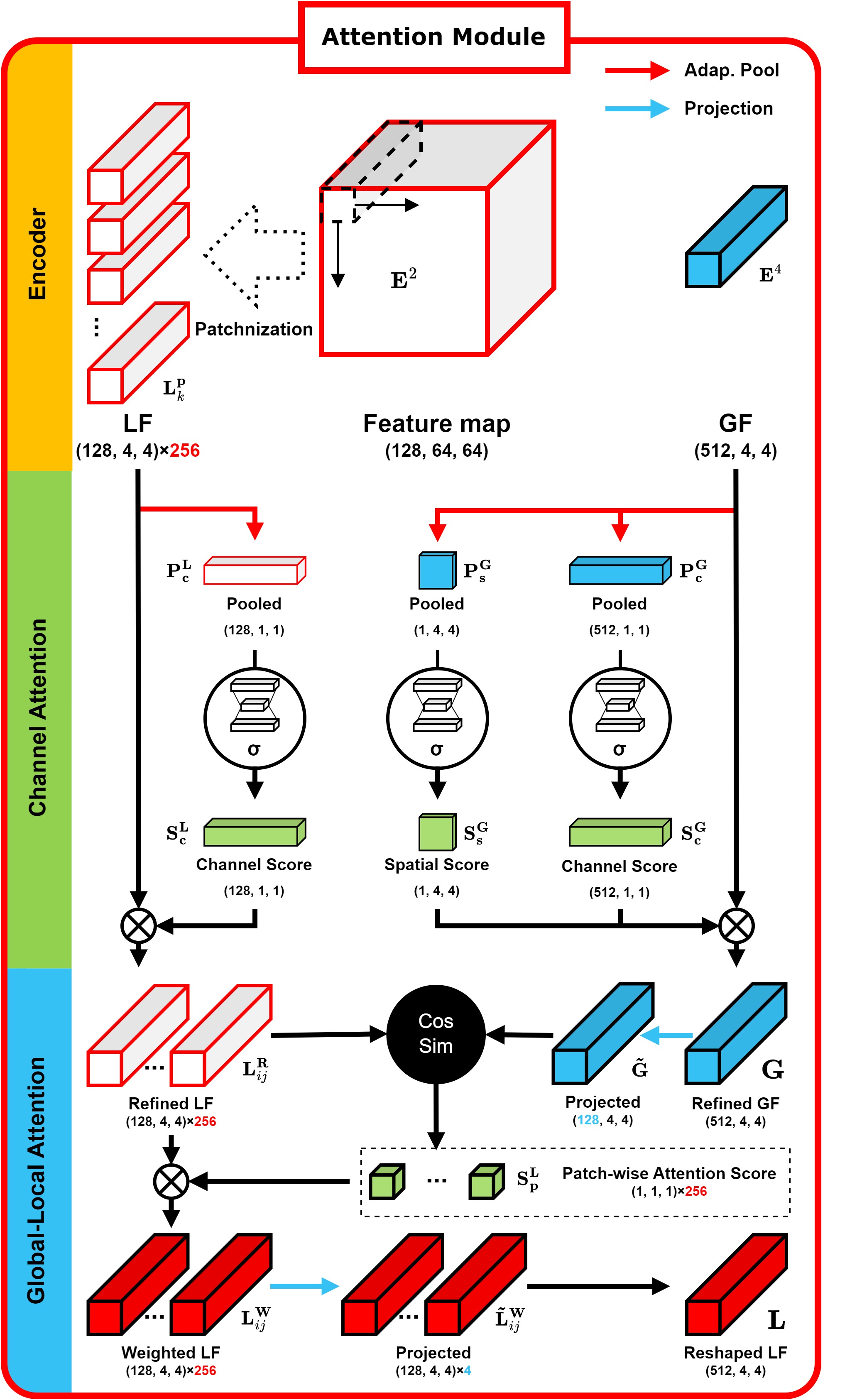}}
        \caption{
    Global-Local Attention Module. The module recalibrates local features $\mathbf{E}^2$ and global features $\mathbf{E}^4$ to enhance motion estimation. 
    Local feature blocks are extracted from $\mathbf{E}^2$ and recalibrated using channel attention, resulting in refined local feature blocks $\mathbf{R}_{k}$. 
    Global features are derived from $\mathbf{E}^4$ through spatial and channel attention, yielding $\mathbf{G}$, which captures semantic information across the entire image. 
    Each local feature block is weighted based on its similarity to the recalibrated global feature $\mathbf{G}$, and weighted local feature blocks $\mathbf{L_k}$ are projected to aggregate local information.
    The reshaped local feature $\mathbf{L}$ and global feature $\mathbf{G}$ serve as the final representations, which are input to the motion estimator.
    }
        \label{fig:GL_ATT}
\end{figure}

In recent advancements within the fields of natural image and language processing, numerous attention mechanisms have been developed to capture dependencies between large-scale (global) and small-scale (local) contexts \cite{huang_2021_gloria,li_2017_GLA_1,le_2022_GLA_2}. Drawing inspiration from these mechanisms, we designed a global-local attention module as a self-attention mechanism. This module highlights local features (semantics in specific regions) based on global features that summarize semantic information across the entire image, allowing the motion estimator to leverage this information efficiently.  

As illustrated in Fig.~\ref{fig:GL_ATT}, the local feature $\mathbf{E}_{i}^2 \in \mathbb{R}^{128 \times 64 \times 64}$, obtained by passing B-mode frames through a few encoding layers, reduces the spatial domain from $128 \times 128$ to $64 \times 64$ while preserving local features of the B-mode domain. 
This feature is then used to derive multiple local feature blocks.

First, both global and local features are recalibrated using standard attention mechanism. For the local feature blocks, the feature map  $\mathbf{E}_{i}^2$ is divided into 256 local feature blocks, 
$\{\mathbf{E}^2_k\}_{k=1}^{256}$, through a patchnization, where $\mathbf{E}^2_k \in  \mathbb{R}^{128 \times 4 \times 4}$.   
 Then, the pooled feature map is calculated as $\mathbf{P_c^L}=\Gamma(\mathbf{E}^2)\in\mathbb{R}^{128}$, where $\Gamma(\cdot)$ denotes an adaptive pooling operation. The channel attention score is then obtained as:
\begin{align} 
        \mathbf{S_c^L} = \sigma(\mathbf{W}^\mathbf{L}_{\mathbf{c}2}\mathbf{W}^\mathbf{L}_{\mathbf{c}1}\mathbf{P_c^L}) \in \mathbb{R}^{128},
\end{align} 
where $\mathbf{W}^\mathbf{L}_{\mathbf{c}2} \in \mathbb{R}^{128\times 8}$ and $\mathbf{W}^\mathbf{L}_{\mathbf{c}1} \in \mathbb{R}^{8\times 128}$ are the weights of a multi-layer perceptron (MLP), and $\sigma(\cdot)$ is the sigmoid function. 
The recalibration of each local patch block $\mathbf{E}^2_k$ is then conducted using the channel attention score:
\begin{align}
        \mathbf{R}_k = \mathbf{S_c^L} \otimes \mathbf{E}^2_k \in \mathbb{R}^{128 \times 4 \times 4},
\end{align}
where $\otimes$ denote hadamard product with broadcasting. 
The resulting recalibrated feature blocks $\mathbf{R}_k$ capture the refined local information.

Meanwhile, the global feature $\mathbf{E}^4 \in \mathbb{R}^{512 \times 4 \times 4}$, generated by passing B-mode frames through deeper encoding layers, spatially condenses information to store comprehensive global semantic features. 
The channel attention scores for global feature are obtained as:
\begin{align*} 
        \mathbf{S_c^G} =&\; \sigma(\mathbf{W}^\mathbf{G}_{\mathbf{c}2}\mathbf{W}^\mathbf{G}_{\mathbf{c}1}\mathbf{P_c^G}) \in \mathbb{R}^{512}, 
\end{align*}
where $\mathbf{W}^\mathbf{G}_{\mathbf{c}2} \in \mathbb{R}^{512\times 32}$ and $\mathbf{W}^\mathbf{G}_{\mathbf{c}1} \in \mathbb{R}^{32\times 512}$ are the weights of a MLP, and $\mathbf{P_s^G}=\Gamma(\mathbf{E}^4)\in\mathbb{R}^{512}$ is a pooled feature derived from $\mathbf{E}^4$.
Next, the spatial attention score for global feature are obtained as:
\begin{align*} 
        \mathbf{S_s^G} =&\; \sigma(\mathbf{W^G_s}*\mathbf{P^G_s}) \in \mathbb{R}^{4 \times 4},
\end{align*}
where $\mathbf{W^G_s}\in\mathbb{R}^{2\times 1\times 1}$ is the kernel of a $1\times 1$ convolutional layer, and $\mathbf{P^G_s}\in\mathbb{R}^{2\times 4\times 4}$ is the concatenation of max-pooled and average-pooled features. Here, $*$ denotes the convolution operation.
Both spatial and channel attention are then applied parallelly as:
\begin{align*} 
        \mathbf{G} =&\; \mathbf{S_s^G} \otimes \mathbf{S_c^G} \otimes \mathbf{E}^4 \in \mathbb{R}^{512 \times 4 \times 4}, 
\end{align*} 
where $\mathbf{G}$ represents the recalibrated global feature.

To prepare for the global-local attention operation, the global feature $\mathbf{G}$ is projected into $\tilde{\mathbf{G}} \in \mathbb{R}^{128 \times 4 \times 4}$ using a $1$$\times$$1$ convolution to match the channel dimensions. Each recalibrated local feature block $\mathbf{R}_{k}$ is then weighted based on its similarity to the global feature using cosine similarity:
\begin{align}
        \mathbf{L}_{k} = \Phi(\mathbf{R}_{k}, \tilde{\mathbf{G}}) \otimes \mathbf{R}_{k},
\end{align}
where $\Phi(\cdot, \cdot) \in \mathbb{R}$ is the cosine similarity function, emphasizing regions critical for motion estimation.  

The weighted local feature blocks $\{\mathbf{L}_{k}\}_{k=1}^{256}$ are projected to $\tilde{\mathbf{L}} \in \mathbb{R}^{128 \times 4 \times 4 \times 4}$ for aggregation of the local information. Finally, the reshaped local feature $\mathbf{L} \in \mathbb{R}^{512 \times 4 \times 4}$ is obtained from $\tilde{\mathbf{L}}$. The outputs of the global-local attention ($\mathbf{L}$ and $\mathbf{G}$) are then fed into the motion estimators.

\subsection{Loss Functions}  
\subsubsection{Motion-based Mean Absolute Error (MMAE)}  
There is a class imbalance among the elements of motion, as cases with high amplitudes (fast motion) are significantly less frequent. Moreover, estimating fast motion poses a greater challenge than estimating slow motion, primarily because the correlation between adjacent frames diminishes. To address this issue, we propose the motion-based mean absolute error (MMAE), a type of weighted metric:  
\begin{align}
    \label{eq:MMAE}
    L_{\text{MMAE}} = 
    \frac{1}{6(s+1)}
    \sum_{i=n}^{n+s} \sum_{k=1}^{6} 
    \mathbf{w}_i \left| \Delta\boldsymbol{\theta}_{i}^{k} - \hat{\Delta\boldsymbol{\theta}_{i}^{k}} \right|,
\end{align}
where $s+1$ is the length of the B-mode sequence, and $\mathbf{w}_i=\left| \Delta\boldsymbol{\theta}_i \right| + \varepsilon$ is a weighting vector defined by the motion vector.
The error for fast motions is weighted more heavily than for slow motions, as $\mathbf{w}_i$ increases for fast motions. Here, $\varepsilon$ is a smoothing factor to reduce the effect of the weighting.  

\subsubsection{Correlation Loss}  
To assist in model supervision, we adapted correlation loss \cite{guo_2020_DCL}, which is defined as:  
\begin{align}
    \label{eq:Loss_Corr}
    L_{\text{Corr}} = 
    \frac{1}{6}
    \sum_{k=1}^{6}
    \left(1 - \Phi \left(
        \left\{\Delta\boldsymbol{\theta}_i^{k}\right\}_{i=n}^{n+s}, 
        \left\{\hat{\Delta\boldsymbol{\theta}_i^{k}}\right\}_{i=n}^{n+s}\right)
    \right),
\end{align}
where $\Phi(\cdot, \cdot)$ denotes the cosine similarity function.  

The correlation loss can measure the error without being affected by scale. Furthermore, it imposes a stronger penalty for incorrectly estimated directions, thereby enhancing the model's stability in capturing accurate motion.  

\subsubsection{Margin Triplet Loss}  
Although each scan can have different appearances in B-mode images due to varying scan protocols or anatomical variations, similar motions can still be found among scan frames. To further assist model convergence, we employed contrastive learning with a margin triplet loss \cite{Schroff_2015_Triplet}:  
\begin{align}
    \label{eq:Loss_Triplet}
    L_{\text{Triplet}} = 
    \max(0, \text{dist}(\mathbf{F}_a, \mathbf{F}_p) - \text{dist}(\mathbf{F}_a, \mathbf{F}_n)),
\end{align}
where $\mathbf{F}_a$ is an anchor feature map serving as a center criterion. To determine the positive and negative feature maps, $\mathbf{F}_p$ and $\mathbf{F}_n$, which correspond to the near and far samples, the cosine similarity between labels is used.  

The triplet loss $L_{\text{Triplet}}$ encourages the model to contrast the feature maps in latent space, facilitating their convergence and improving robustness.  

\subsubsection{Final Loss Function}  
The final loss function is a linear combination of the individual loss components:  
\begin{align}
    \label{eq:Loss_Final}
    L_{\text{Final}} = 
    \alpha_1 L_{\text{MMAE}} + \alpha_2 L_{\text{Corr}} + \alpha_3 L_{\text{Triplet}},
\end{align}
where $\alpha_1$, $\alpha_2$, and $\alpha_3$ are positive real values used to balance the magnitude of each loss function.  

\section{Experimental Results}
\subsection{Meterials}
\begin{figure}[t]
        \captionsetup{labelfont=bf}
        \centerline{\includegraphics[width=\columnwidth]{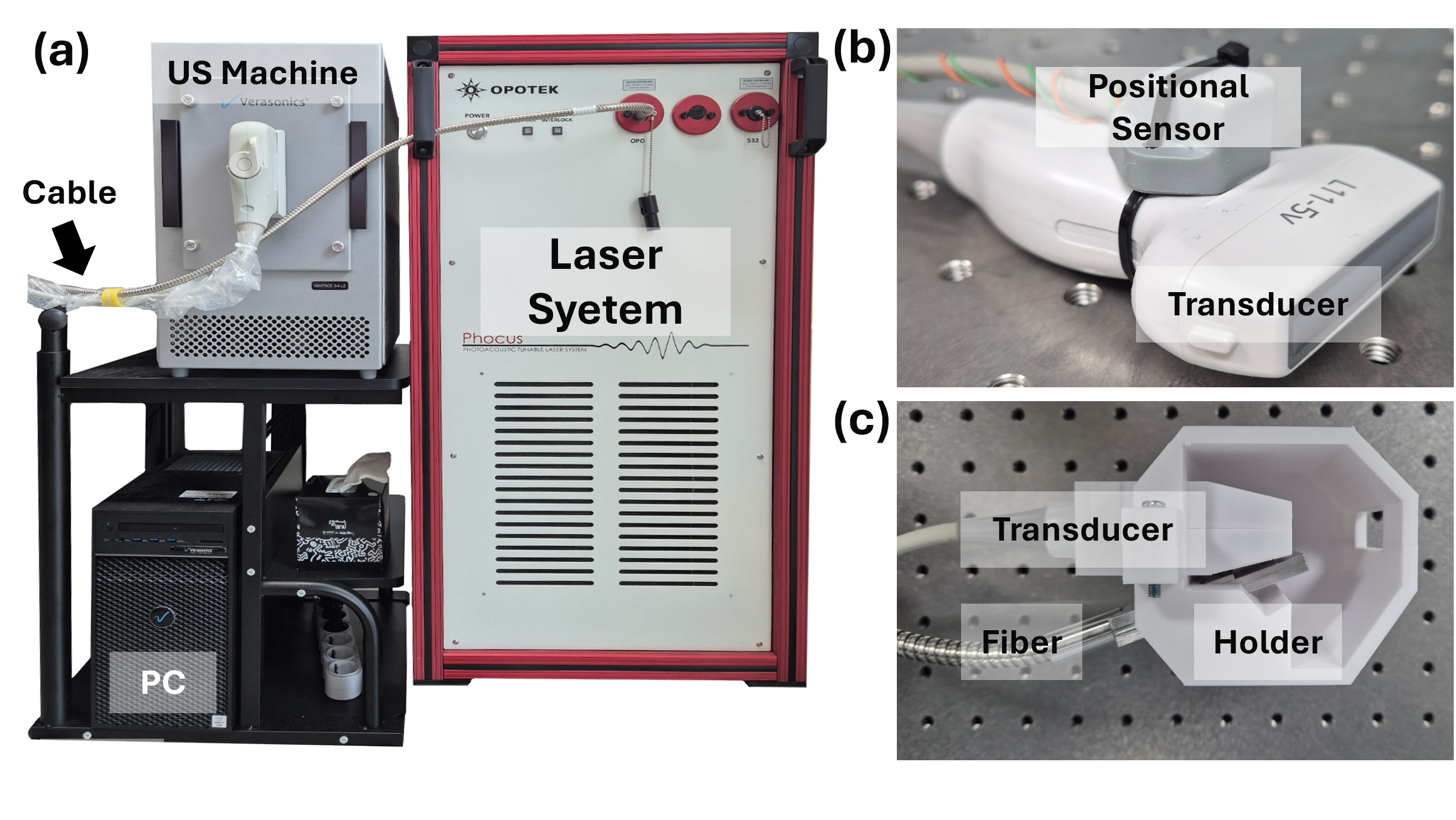}}
\vspace{-10pt}
        \caption{Experimental setup for PAUS data acquisition and visualized results. 
    (a) US machine connected to the laser system. 
    (b) and (c) Transducer setup. 
    }
\vspace{-10pt}
        \label{fig:US_setup}
\end{figure}

\subsubsection{Dataset 1}
We used a programmable US machine (Verasonics, Vantage System 64LE) equipped 
with a 1D array transducer (L11-5v) to acquire US data,  
and the scan motion was tracked using a mounted electromagnetic sensor (Polhemus) 
at an acquisition rate of 60 Hz (Fig.~\ref{fig:US_setup}). 
The B-mode images were reconstructed using 31 different angle plane waves 
to enhance their quality and have dimensions of 256$\times$256 pixels, 
with a pixel resolution of 0.1484$\times$0.1484 mm. 
The image frame rate was set at 20 Hz. 
We collected 54 scans, primarily from the forearms of 9 individuals, 
and randomly divided them into 30, 6, and 18 scans 
for the training, validation, and testing sets, respectively, 
ensuring that the subjects were distinct across sets. 
Each scan consisted of 900-1000 frames, 
capturing the entire length of the forearm through an arbitrary S-shaped trajectory. 
This scanning procedure was conducted three times for each forearm. 
The local Institutional Review Board (Pusan National Univ. IRB, 2023\_74\_HR) 
granted approval for the acquisition and use of the data.

\subsubsection{Dataset 2}
To construct 3D vascular structures, 
we acquired US Doppler data using the same US acquisition system. 
Iimmediately following the emission of 31 plane waves for a B-mode image frame, 
20 plane waves with a pulse repetition frequency (PRF) of 3000 kHz were transmitted 
for a Power Doppler (PD) image frame. 
To minimize clutter signals, a filter based on singular value decomposition (SVD) was utilized.
Additionally, photoacoustic (PA) data were collected to construct 3D vascular structures. 
As shown in Fig.~\ref{fig:US_setup} (a), 
an Q-switched Nd:YAG lasers system (OPOTEK Phocus Mobile) was integrated with the US machine 
to enable a real-time PAUS imaging. 
Specifically, the US system received a flash lamp trigger signal from the laser system, 
waited for a specified delay time (optical build-up time for the laser), 
and then commenced data acquisition precisely when the laser fired. 
The laser was delivered via fibers arranged 
on one side of the linear array US transducer (Fig.~\ref{fig:US_setup} (c)).
Subsequent to the 31 plane waves used for a B-mode image frame, 
a laser with a wavelength of 750 nm was employed to insonify for a PA image frame. 
PA images were reconstructed using a standard delay-and-sum (DAS) method. 

\subsubsection{Dataset 3}
Our dataset may contain inherent biases that could positively influence our model's performance. To validate the model's performance in different environments and mitigate the effects of such biases, we utilized a publicly available dataset \cite{li_2023_EfficientNet}. This dataset comprises transverse and longitudinal scans of the forearm with diverse-shaped trajectories, collected from 19 subjects, resulting in a total of 19$\times$12 scans. For each subject, both forearms were scanned using three distinct trajectory shapes (S, C, and L) in two orientations: parallel and perpendicular to the forearm, yielding 12 scans per subject.

The ultrasound images were acquired using an Ultrasonix machine (BK, Europe) operating at 20 Hz with a convex transducer (4DC7-3/40). The probe motion was tracked by an NDI Polaris Vicra (Northern Digital Inc., Canada). The B-mode images were processed with a median level of speckle reduction.
Each scan consists of 36 to 430 frames with a resolution of 480$\times$640 pixels, and the probe travel distance ranges approximately between 100 and 200 mm. The dataset was randomly divided into training, validation, and testing subsets based on subjects, with a ratio of 10:4:5.

Since images captured using the convex probe have a sector shape with background regions outside the sector, we extracted a square region of interest (320$\times$320 pixels) from the central foreground region to focus on relevant areas. To handle the increased input size, the batch size was reduced to 12.

\subsection{Implementation Details}
For the comparative experiments, we selected models currently recongnized for motion estimation, 
including CNN \cite{prevost_2018_CNN}, 
DC\textsuperscript{2}-Net \cite{guo_2022_DC2}, 
EfficientNet \cite{li_2023_EfficientNet}, 
and LSTM \cite{luo_2023_RecO}. 
We made minor adjustments to the structure and hyperparameters of each model 
to tailor them to our task.
During the training phase, 
an arbitrary sequence of length $s$ is sampled from each person in every epoch. 
This  method requires a relatively large number of epochs (20,000) to ensure model convergence. 

The Adam optimizer was employed with an initial learning rate of 1e-5, which was reduced by a factor of 0.8 every 100 epochs to facilitate the convergence process. Training was performed on an NVIDIA RTX 4090 GPU (24 GB) with a batch size of 14. Under these settings, the proposed MoGlo-Net required 4.7 hours to train and achieved a processing speed of 95 frames per second (FPS) with a batch size of 32 during the inference phase. Additionally, reconstructing a 3D US volume consisting of 950 frames took 1.8076 seconds.

All experiments were conducted under identical conditions, including sequence length, dataset splitting, and training parameters, to ensure a fair comparison among the models.

\subsection{Evaluation Metrics}
For the 6-dimensional output, we utilized the relative Average Error (\textbf{rAE}; mm, °) and accumulated Average Error (\textbf{aAE}; mm, °) to evaluate $\hat{\Delta\boldsymbol{\theta}_{i}}$ and $\hat{\boldsymbol{\theta}_{i}}$. The rAE measures the relative motion estimation accuracy, while the aAE quantifies the accumulated motion error over the scan trajectory. For the reconstructed 3D ultrasound volumes, we measured the Euclidean distance of the grid points between the true and predicted frames. The relative Frame Error (\textbf{rFE}; mm) reflects the relative displacements between frames, whereas the accumulated Frame Error (\textbf{aFE}; mm) accounts for the cumulative displacements along the scan path. Additionally, we used Correlation (\textbf{Corr}), defined as cosine similarity, to measure the underlying trends of reconstructed trajectories in 3D Euclidean space. The Final Drift (\textbf{FD}; mm), which represents the aFE of the final frame, increases proportionally with the scan length. To normalize FD, we employed the Final Drift Rate (\textbf{FDR}; mm), calculated by dividing FD by the total length of the scan trajectory. FDR serves as a subsidiary metric, as FD typically exhibits relatively large deviations in S-shaped scans.

\subsection{Results using Dataset 1}

\subsubsection{Quantitative Results}
As shown in Table~\ref{table:Model_Comparison}, the DCL shows better performance than conventional CNN, 
with further improvement achieved through contrastve learning in DC\textsubscript{2}. 
Efficient yields more precise results by leveraging sequencial information, 
while LSTM provides more reliable performance by incorporating motion-based constraint. 
The MoGLo (ours) achieved the best performance across all metrics among the comparison models.

\renewcommand{\arraystretch}{1.2}
\begin{table}[htbp]
        \captionsetup{labelfont=bf}
        \caption{Model Performance - In-House Dataset}
		\small
		\small
        \begin{tabularx}{\columnwidth}{ p{0.17\columnwidth} *{6}{X} }
                \toprule
                Models & 
                rAE & aAE & rFE & aFE & Corr & FDR \\ 
                \midrule
                CNN\cite{prevost_2018_CNN} & 
                0.1946 & 17.9297 & 0.5045 & 40.5307 & 0.5282 & 32.7253 \\
                DCL\cite{guo_2020_DCL} & 
                0.1272 & 14.6046 & 0.3463 & 39.4931 & 0.8464 & 29.5583 \\
                DC\textsuperscript{2}\cite{guo_2022_DC2} & 
                0.1361 & 11.2227 & 0.3584 & 29.8875 & 0.8775 & 21.3799 \\ 
                Efficient\cite{li_2023_EfficientNet} & 
                0.1177 & 10.5001 & 0.3214 & 25.4893 & 0.7999 & 21.8983 \\ 
                LSTM\cite{luo_2023_RecO} & 
                0.1159 & 8.8340 & 0.3212 & 21.9479 & 0.8826 & 15.5101 \\ 
                \textbf{MoGLo} & 
                \textbf{0.1047} & \textbf{6.9254} & \textbf{0.2917} & \textbf{16.1944} & \textbf{0.9217} &\textbf{12.0577} \\ 
                \bottomrule
        \end{tabularx}
        \label{table:Model_Comparison}
\end{table}

\subsubsection{Qualitative Results}
\begin{figure}[ht]
        \captionsetup{labelfont=bf}
        \centerline{\includegraphics[width=\columnwidth]{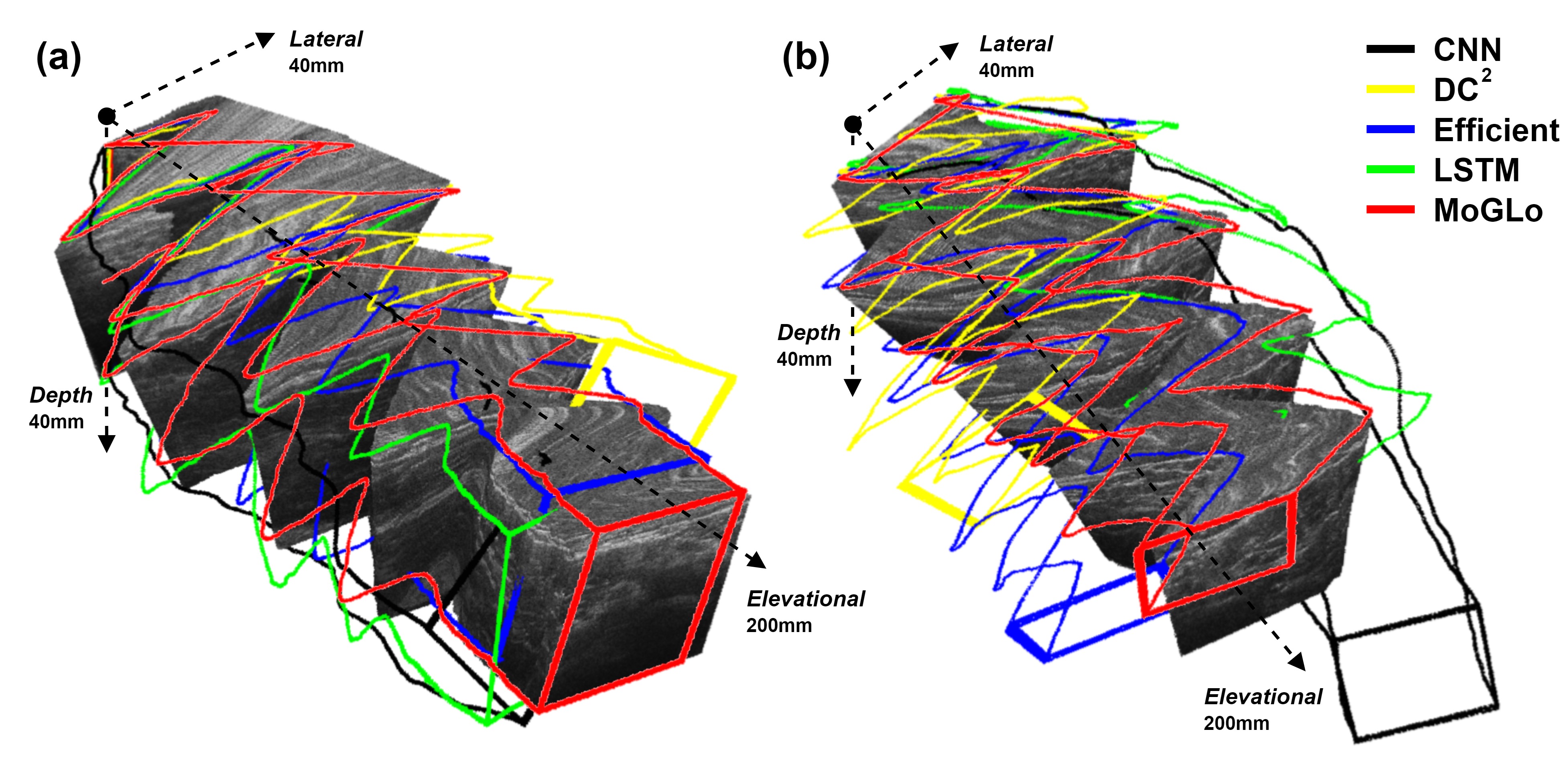}}
       \caption{Two 3D reconstruction cases using US B-mode acquisitions. Each case is shown from two different directional views. The 3D ground-truth image is constructed by stacking 2D B-mode images using ground-truth positions. The differently colored outlined 3D figures (no filled) are constructed using estimated positions from various deep learning models to compare their trajectories with the ground truth.}
        \label{fig:US3D_B}
\vspace{-5pt}
\end{figure}

Consequently, the final 3D US volume can be reconstructed using the estimated motion and B-mode images. 
As illustrated in Fig.~\ref{fig:US3D_B} (a), 
CNN exhibits pronounced distortion, 
while DC\textsuperscript{2} and Efficient show noticeable improvements over CNN. 
However, both models display considerable FD due to their limited generalization capabilities. 
Among the comparison groups, LSTM demonstrates the most reliable results in terms of both 
trajectory similarity and FD. 
Models that utilize correlation volume and global-local attention 
tend to futher improve the quality of 3D US volumes, 
with the proposed MoGLo-Net showcasing the most accurate results, 
closely resembling the ground truth.

\subsubsection{Ablation Study 1}
\begin{table}[htbp]
        \captionsetup{labelfont=bf}
        \caption{Ablation Study 1 - Model Components}
		\small
        \begin{tabularx}{\columnwidth}{ p{0.01\columnwidth} p{0.01\columnwidth} p{0.055\columnwidth} *{6}{X}}
                \toprule
                G & C & M & 
                rAE & aAE & rFE & aFE & Corr & FDR\\ 
                \midrule
                $\times$ & $\times$ & $\times$ & 
                0.1142 & 8.6565 & 0.3127 & 20.1755 & 0.8723 & 17.4204 \\ 
                \checkmark & $\times$ & $\times$ & 
                0.1091 & 8.4476 & 0.3002 & 18.8602 & 0.9008 & 14.6913 \\ 
                $\times$ & \checkmark & $\times$ & 
                0.1109 & 8.4749 & 0.3074 & 20.3687 & 0.8806 & 14.8185 \\ 
                $\times$ & $\times$ & \checkmark & 
                0.1123 & 8.2902 & 0.3081 & 19.8897 & 0.8900 & 15.7800 \\ 
                \checkmark & \checkmark & $\times$ & 
                0.1094 & 7.2300 & 0.3018 & 16.7405 & 0.9060 & 14.1837 \\ 
                \checkmark & $\times$ & \checkmark & 
                0.1083 & 7.0348 & 0.2981 & 17.0640 & 0.9002 & 13.5590 \\ 
                $\times$& \checkmark & \checkmark & 
                0.1087 & 8.0705 & 0.3019 & 19.1706 & 0.8998 & 15.1482 \\ 
                \checkmark & \checkmark & \checkmark & 
                \textbf{0.1047} & \textbf{6.9254} & \textbf{0.2917} & \textbf{16.1944} & \textbf{0.9217} & \textbf{12.0577} \\ 
                \bottomrule
        \end{tabularx}
        \label{table:Ablation_Study1}
\end{table}

We created a minor version of MoGLo-Net by removing the global-local attention module (G), correlation operation (C), or motion-based MAE (M) to further validate the contribution of each component in MoGLo-Net. As shown in Table~\ref{table:Ablation_Study1}, the global-local attention significantly improved overall metrics, particularly in accumulative errors, as it effectively highlights critical regions for motion estimation. Employing the correlation operation resulted in partial metric improvements, as it provides comprehensive but unrefined correlation information between frames. Finally, the inclusion of MMAE led to improvements, especially in accumulative errors, by emphasizing fast motion, which is sparsely distributed. By leveraging all components, MoGLo-Net exhibits significantly less distortion compared to its minor version.

\subsubsection{Ablation Study 2}
\begin{table}[htbp]
        \captionsetup{labelfont=bf}
        \caption{Ablation Study 2 - Input Data Types}
		\small
        \begin{tabularx}{\columnwidth}{ p{0.17\columnwidth} *{6}{X} }
                \toprule
                Models & 
                rAE & aAE & rFE & aFE & Corr & FDR \\ 
                \midrule
                Log & 
                0.1127 & 9.7552 & 0.3105 & 25.5322 & 0.8790 & 18.9047 \\
                IQ & 
                0.1115 & 7.3152 & 0.3157 & 18.8984 & 0.8925 & 13.5999 \\
                SP & 
                0.1088 & 7.8099 & 0.2988 & 18.9144 & 0.9021 & 14.8614 \\
                \textbf{MoGLo} & 
                \textbf{0.1047} & \textbf{6.9254} & \textbf{0.2917} & \textbf{16.1944} & \textbf{0.9217} & \textbf{12.0577} \\ 
                \bottomrule
        \end{tabularx}
        \label{table:Ablation_Study2}
\end{table}

We also examined the effect of input data type by leveraging the programmable ultrasound machine. We designed three settings, which are summarized in Table~\ref{table:Ablation_Study2}: 
1) Log: inputting rawer B-mode images, referring to the images before log-compression for contrast enhancement; 
2) IQ: inputting in-phase and quadrature (IQ) data as well as B-mode images. To utilize the IQ data, we applied log compression to its amplitude and concatenated it with the B-mode; 
and 3) SP: inputting B-mode images after applying noise reduction to mitigate speckle noise using the Lee Filter \cite{lee_1980_LeeFilter}. 

As shown in Table~\ref{table:Ablation_Study2}, when raw B-mode images were used, overall performance decreased due to poor contrast in dark regions, which occupy the majority of the image. Adding IQ data did not yield improvement, as it increased the complexity of the input features without providing additional cues for motion estimation. Lastly, using denoised B-mode images reduced overall performance, as speckle patterns are strongly correlated with motion vectors.

\subsection{Results using Dataset 2}

\begin{figure}[ht]
    \captionsetup{labelfont=bf}
    \centerline{\includegraphics[width=\columnwidth]{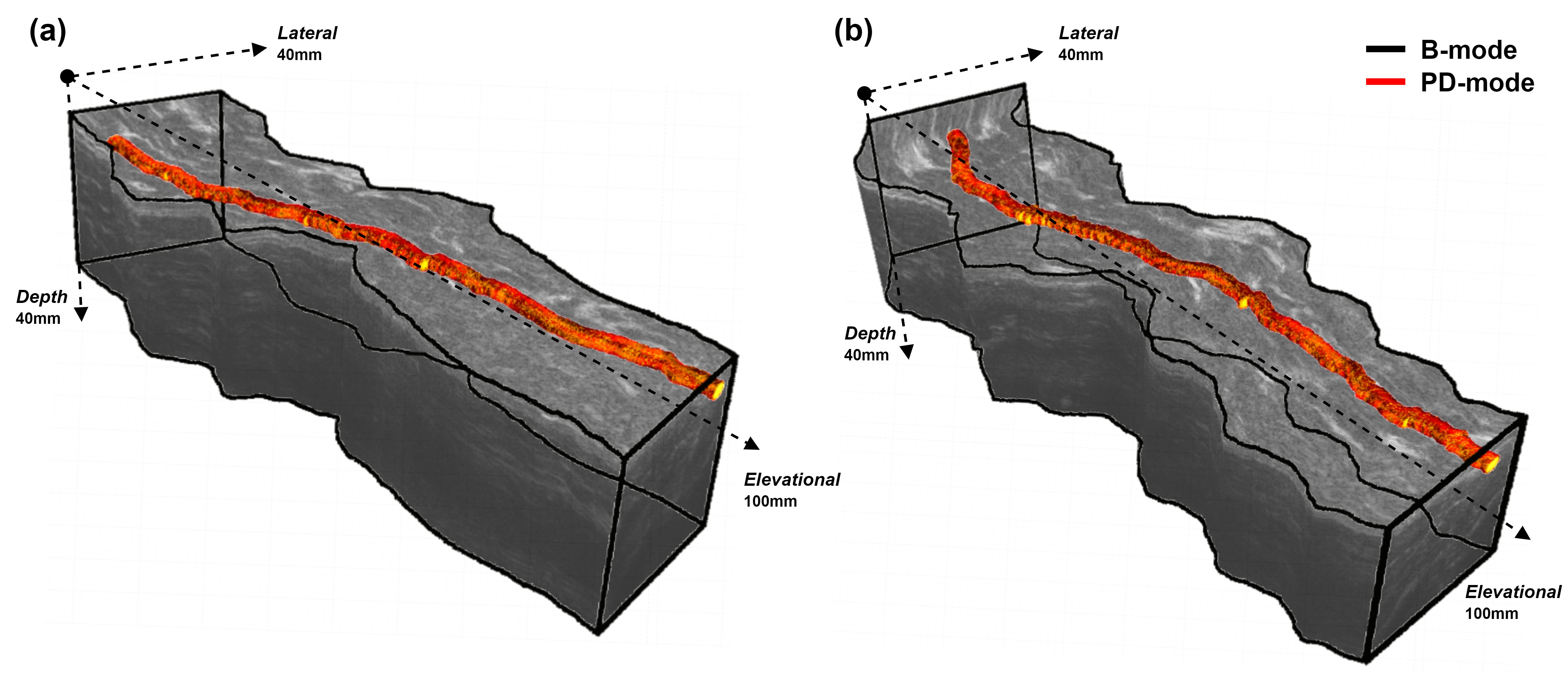}}
    \caption{Two 3D vascular (radial artery in a forearm) reconstruction cases using US PD acquisitions. Each 3D vessel is visualized by superimposing B-mode imaging to check the scan motion trajectory.}
    \label{fig:US3D_D}
\end{figure}

The benefit of motion tracking extends the clinical applications beyond the compilation of B-mode images. In this study, we adapted the acquisition sequence in the ultrasound system to obtain not only B-mode images but also Power Doppler (PD) mode images, which are specialized for visualizing vessels. B-mode images were utilized for motion estimation through MoGLo-Net, and both the motion estimates and PD images were employed to construct 3D vessels. 

Fig.~\ref{fig:US3D_D} illustrates examples of reconstructed 3D vessels (radial artery) in a forearm. Notably, each vessel appears almost straight and closely resembles the natural anatomical shape, even though the scan motion trajectory was wavy.

\begin{figure}[ht]
        \captionsetup{labelfont=bf}
        \centerline{\includegraphics[width=\columnwidth]{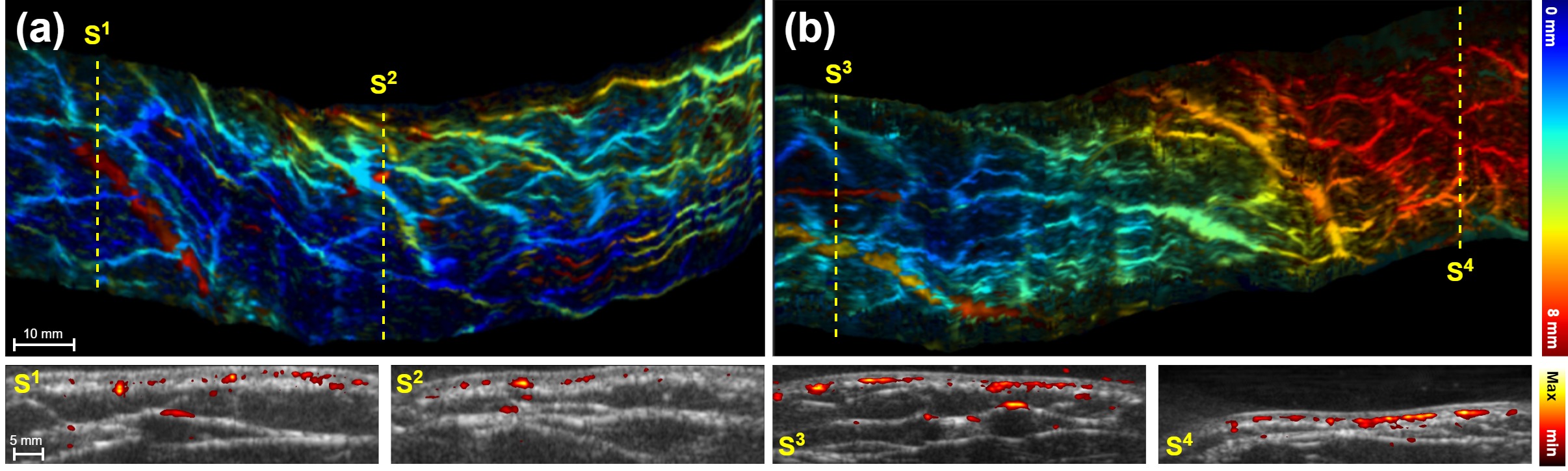}}
        \caption{Two 3D vascular reconstruction cases using PAUS acquisitions. Each is visualized using maximum amplitude projection (mAP) according to the depth, alongside corresponding cross-sectional B-mode images.}
        \label{fig:US3D_P}
\vspace{-5pt}
\end{figure}

Furthermore, we set up the PAUS system to acquire both B-mode and photoacoustic (PA) images, enabling the reconstruction of finer vessels (in a forearm) with high contrast. 
Fig.~\ref{fig:US3D_P} illustrates the PA imaging results using maximum amplitude projection (mAP) according to the depth.

\subsection{Results using Dataset 3}
\renewcommand{\arraystretch}{1.2}
\begin{table}[htbp]
        \captionsetup{labelfont=bf}
        \caption{Model Performance - Open Dataset}
		\small
        \begin{tabularx}{\columnwidth}{ p{0.17\columnwidth} *{6}{X} }
                \toprule
                Models & 
                rAE & aAE & rFE & aFE & Corr & FDR \\ 
                \midrule
                CNN\cite{prevost_2018_CNN} & 
                0.2183 & 9.9857 & 0.5866 & 29.8432 & 0.7749 & 37.1222 \\
                DCL\cite{guo_2020_DCL} & 
                0.1988 & 9.1184 & 0.5281 & 27.2999 & 0.8131 & 34.9391 \\
                DC\textsuperscript{2}\cite{guo_2022_DC2} & 
                0.1947 & 8.8892 & 0.5146 & 27.1966 & 0.8206 & 35.1307 \\ 
                Efficient\cite{li_2023_EfficientNet} & 
                0.1853 & 7.5940 & 0.4495 & 19.5572 & 0.8672 & 27.0905 \\ 
                LSTM\cite{luo_2023_RecO} & 
                0.1815 & 7.1584 & 0.4437 & 19.5029 & 0.8751 & 25.1588 \\ 
                \textbf{MoGLo} & 
                \textbf{0.1745} & \textbf{6.7183} & \textbf{0.4183} & \textbf{17.8089} & \textbf{0.8940} &\textbf{22.6942} \\ 
                \bottomrule
        \end{tabularx}
\vspace{-5pt}
        \label{table:Model_Comparison_2}
\end{table}

\begin{figure}[h]
	\captionsetup{labelfont=bf}
	\centerline{\includegraphics[width=\columnwidth]{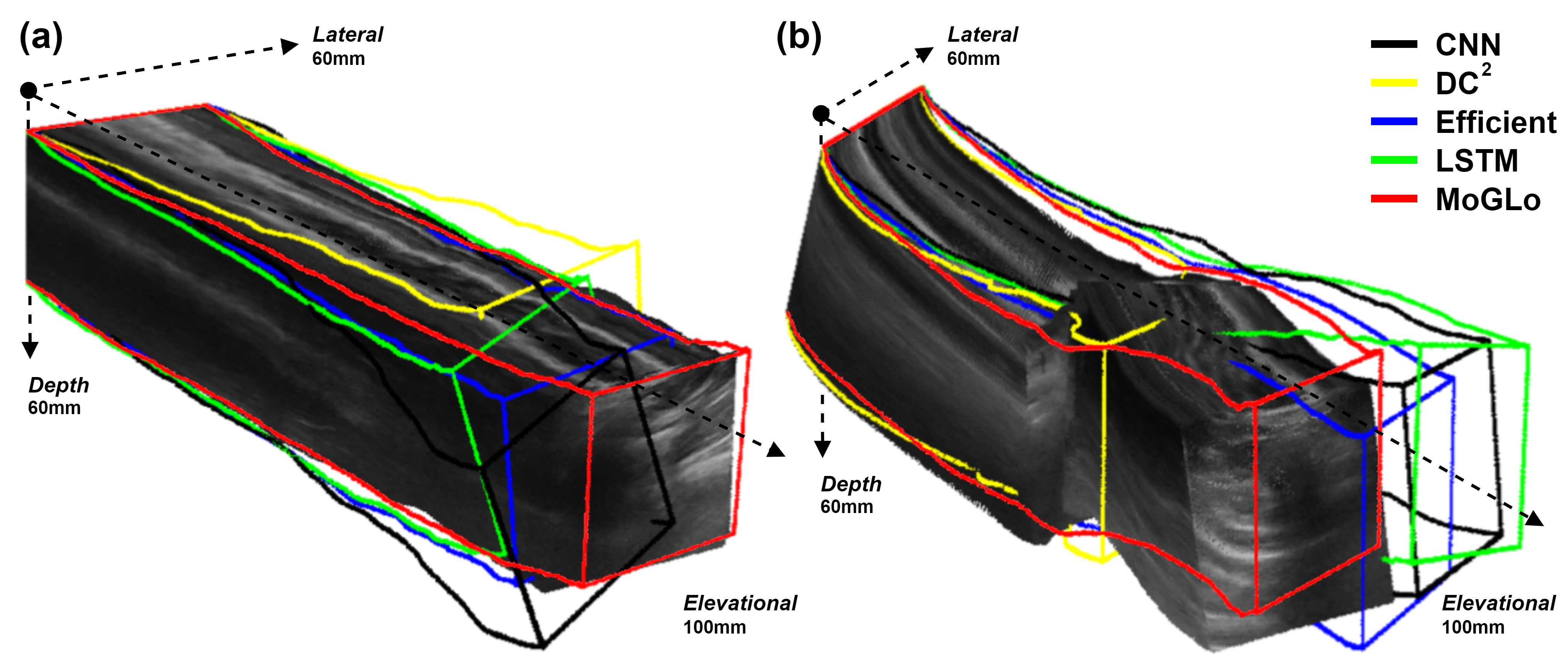}}
   \caption{Two 3D reconstruction cases using publicly open dataset. The 3D ground-truth image is constructed by stacking 2D B-mode images using ground-truth positions. The differently colored outlined 3D figures (no filled) are constructed using estimated positions from various deep learning models to compare their trajectories with the ground truth.}
	\label{fig:US3D_B_2}
\end{figure}

As shown in Table \ref{table:Model_Comparison_2}, performance metrics varied across models due to differences in dataset features, such as hardware setup, scan protocol, and image processing. However, the overall performance trends remained consistent across models.

The consistency in the performance order observed between the in-house dataset and the open dataset strengthens the reliability of the experimental findings. This consistency indicates that the observed performance is not specific to a particular dataset but instead reflects inherent characteristics of the models. Fig.~\ref{fig:US3D_B_2} illustrates 3D trajectories using ground-truth and estimated positions.

\section{Discussion and Conclusion}
The primary advantage of handheld PAUS systems is the flexibility they offer during scanning. 
However, this comes at the cost of requiring skilled and experienced operators 
with a deep understanding of anatomical 3D structures, 
as standard 1D array transducers provide only 2D images with 
a restricted FoV. 
These challenges can be addressed by reconstructing 3D volumes, 
which allow the visualization of complex 3D structures and 
provide arbitrary cross-sections of 
the RoI. 
Historically, there have been various hardware-based approaches \cite{yen_2000_2D_array_transducer, db_1995_Mechanical_transducer} 
to obtain 3D volumes in both PA and US fields. 

The ideal solution, however, lies in software-only approaches, which are more cost-effective. 
The freehand 3D approach is an extension of the panoramic image mode, 
which has already been commercialized in many ultrasound products. 
Nevertheless, one of the key challenges remains addressing elevational motion (out-of-plane motion), 
where two adjacent frames have relatively low correlation. 
Some groups have introduced the use of IMU sensors attached to the transducer as minimal hardware support to aid motion estimation in software. 
However, bulkier transducers hinder the operator's ability to scan, 
particularly for PAUS systems, which are already large due to the inclusion of laser fibers.

Our proposed model, MoGLo-Net, is optimized for tracking motion without the need for additional sensors. The model employs correlation operations between feature maps from adjacent B-mode images, explicitly utilizing their closeness. Since the correlation tensor captures all correlations between subspaces across sequential images, the model can track not only in-plane motion but also out-of-plane motion.

\begin{figure}[t]
	\captionsetup{labelfont=bf}
	\centerline{\includegraphics[width=\columnwidth]{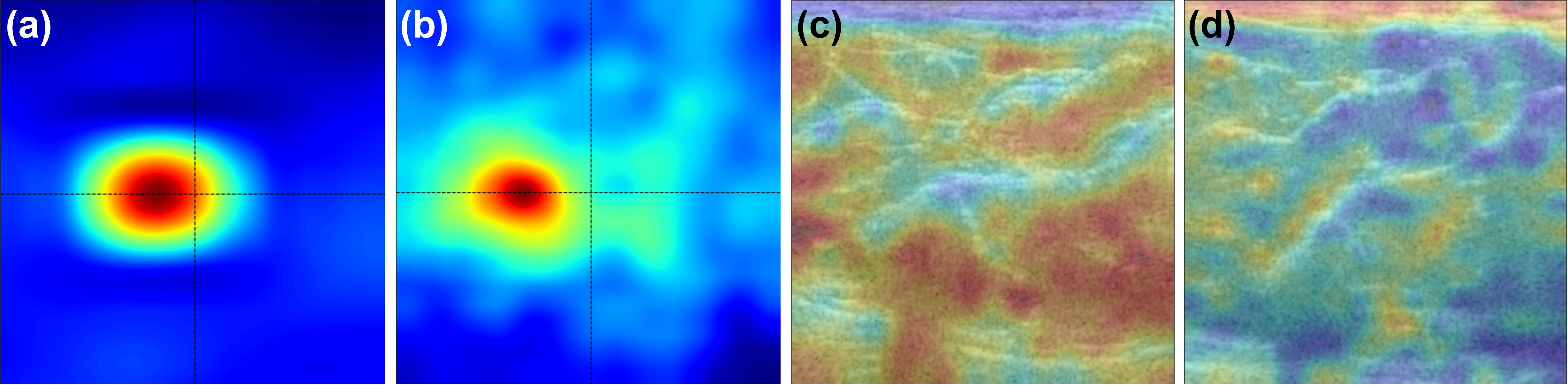}}
	\caption{(a) 2D map obtained by averaging the correlation volume from encoded feature maps over the $z$-axis in Fig.~\ref{fig:Corr}, and (b) corresponding 2D map obtained in the same manner from B-mode images. (c) and (d) Patch-wise attention scores in the global-local attention module overlaid on corresponding B-mode images under conditions of slow and fast elevational motions, respectively.}
\vspace{-10pt}
 \label{fig:Map}
\end{figure}

In additional experiments, we observed that the 3D volume generated from encoded feature maps ($\mathbf{E}^1_i$ and $\mathbf{E}^1_{i+1}$) is more efficient than one generated directly from B-mode images ($\mathbf{B}_i$ and $\mathbf{B}_{i+1}$). Fig.~\ref{fig:Map} (a) shows the 2D map obtained by averaging the correlation volume (from B-mode images) over the $z$-axis in Fig.~\ref{fig:Corr}, while Fig.~\ref{fig:Map} (b) shows the corresponding 2D map obtained in the same manner from encoded feature maps. The feature correlation demonstrates better contrast than B-mode correlation, displaying clear patterns.
The relative value differences over space in the map are expected to aid in estimating in-plane motion, while the overall mean value in the space is expected to aid in estimating out-of-plane motion.

Additionally, the global-local attention module generates local representations of images by learning attention weights that emphasize significant sub-regions for global motion features. This allows the model to effectively highlight sub-regions, such as fully-developed speckle areas, which are critical for motion estimation. 

In extra experiments, we visualized the patch-wise attention scores in the module (Fig.~\ref{fig:GL_ATT}), as shown in Fig.~\ref{fig:Map}. Under conditions of slow elevational motion, adjacent frames share significant overlapping regions, preserving speckle patterns. As a result, the model focused on uniform areas, such as fully-developed speckle regions (Fig.~\ref{fig:Map} (c)). In contrast, under conditions of fast elevational motion, speckle patterns become disrupted due to the larger gap between adjacent frames. Despite this, high-echogenic tissues often maintain consistency along the elevational axis. In such cases, the model highlighted these regions, as illustrated in Fig.~\ref{fig:Map} (d).

Finally, the model utilizes both global and local representations to leverage complementary mutual information, while the LSTM processes features from the B-mode sequences to capture temporal motion patterns. 
The MoGLo-Net demonstrated consistent improvements across all metrics over the state-of-the-art (SOTA) model \cite{luo_2023_RecO}, producing more realistic 3D US volumes.
Moreover, we validated the effect of each component of MoGLo-Net through an ablation experiment by omitting key parts, such as the attention module, correlation volume, and motion-based MAE. In addition, we examined the impact of potential information sources, such as IQ data, raw B-mode images, and speckle patterns. As a result, we found that speckle patterns, which are highly related to scan motion, serve as critical cues for scan motion estimation.

To the best of our knowledge, 
this is the first attempt at free-hand 3D vascular imaging 
using ultrasound Doppler acquisition data or photoacoustic data. 
In this study, we specifically targeted small vessels to emphasize their shape and structure. 
For ultrasound imaging, we chose power-Doppler over color-Doppler image processing 
due to its superior sensitivity in detecting weak blood flow signals in smaller vessels. 
For photoacoustic (PA) imaging, we selected an optimal laser wavelength 
to maximize blood signal detection and reconstructed the images from raw data 
using appropriate hyper-parameters in the DAS 
and filtering procedures.

For small vessels, quantitative parameters such as blood volume and flow velocity 
tend to remain relatively stable over time. 
In contrast, larger vessels exhibit significant fluctuations in these values 
due to pulsatile flow. 
Thus, for accurate quantitative imaging, 
the ideal target and conditions are those where 
the values exhibit minimal temporal variation but may vary spatially. 
This ensures that spatial changes in vascular structure are captured effectively 
without being obscured by temporal variations in blood flow during at least scanning.

The free-hand 3D PAUS technique holds immense clinical potential 
in dignostic imaging and related interventions. 
For example, in the assessment of thyroid nodules, 
hands-free 3D PA imaging could provide detailed visualization of the nodule's vascularity 
and tissue composition in real time. 
PA imaging has the unique ability to highlight abnormal blood vessel growth, 
which is often associated with malignant nodules. 
By integrating 3D reconstructions, clinicians could more accurately differentiate 
between benign and malignant thyroid nodules, 
improving the precision of fine-needle aspiration biopsies 
and reducing the number of unnecessary procedures. 
These applications are the focus of our ongoing studies.

However, in clinical settings, even minor errors can have significant implications, emphasizing the need for prediction models with enhanced precision and accuracy. Furthermore, the methodology proposed in this study exhibits generalizability primarily on forearm scans, necessitating an expansion of the dataset to encompass diverse body parts.

In conclusion, we introduced MoGLo-Net, a novel approach for estimating PAUS scan motion without the need for an additional positional sensor. The scan data encompassed complex motions, including dynamic in-plane movements and unidirectional out-of-plane shifts. These scans often trace relatively long trajectories exceeding 200 mm, which complicates the task as accumulated errors tend to magnify in proportion to the scan length. Thus, achieving generalization performance requires not only minimizing frame-wise errors but also reducing bias and final drift. 
By leveraging specialized supervision methods and deep learning architectures, we have attained superior performance, particularly in terms of accumulated errors. For future work, we aim to enhance this study by integrating anatomical information of tissues or embedding the order of scan sequences to estimate arbitrary motions more accurately. This development promises to further refine motion estimation in ultrasound imaging, potentially leading to more precise and reliable diagnostic tools.

\bibliographystyle{ieeetr}
\bibliography{reference}

\end{document}